\documentclass[aps,pra,reprint,twocolumn,showpacs,floatfix,superscriptaddress]{revtex4-1}

\usepackage{amssymb,amsmath,amstext}                %%   American Physical Society math etc extensions
\usepackage{graphicx}                                               %%   Include figure files                                            
\usepackage{epstopdf}                                               %%   help with eps -> pdf 
\usepackage{color}                                                     %%   change color
\usepackage{bm}                                                        %%   bold math                                    
\usepackage{appendix}                                              %%   formating appendicies
\usepackage[utf8]{inputenc}
\usepackage{latexsym}
\usepackage{xcolor}
\usepackage{braket}
\usepackage{ulem}
\definecolor{lblue} {RGB}{51,71,158}
\usepackage[colorlinks=true,citecolor=blue,linkcolor=blue,urlcolor=lblue]{hyperref}

\begin{document}

\title{Model of level statistics for disordered interacting quantum many-body systems
}% Force line breaks with \\
\author{Piotr Sierant}
\affiliation{Institute of Theoretical Physics, Jagiellonian University in Krakow,  \L{}ojasiewicza 11, 30-348 Krak\'ow, Poland }
\email{piotr.sierant@uj.edu.pl}
\affiliation{ICFO - Institut de Sciences Fotoniques, The Barcelona Institute of Science and Technology, 08860 Castelldefels (Barcelona), Spain}

\author{Jakub Zakrzewski}
\affiliation{Institute of Theoretical Physics, Jagiellonian University in Krakow,  \L{}ojasiewicza 11, 30-348 Krak\'ow, Poland }
\affiliation{Mark Kac Complex
Systems Research Center, Jagiellonian University in Krakow, Krak\'ow,
Poland. }
\email{jakub.zakrzewski@uj.edu.pl}

\date{\today}% It is always \today, today,

%\pacs{03.67.Lx, 42.50.Dv}% PACS, the Physics and Astronomy
                              
\begin{abstract}

We numerically study level statistics of disordered interacting quantum many-body systems.
A two-parameter plasma model which controls  level repulsion exponent $\beta$ and range $h$ of interactions between
eigenvalues is shown to reproduce accurately features of level statistics across the transition from ergodic to many-body 
localized phase. Analysis of higher order spacing ratios indicates that the considered $\beta$-$h$ model accounts 
even for long range spectral correlations and allows to obtain a clear picture of the
flow of level statistics across the transition.
Comparing SFFs of $\beta$-$h$ model and of a system across the ergodic-MBL transition,
we show that the range of effective interactions between eigenvalues $h$ is related to the Thouless time 
which marks the onset of  quantum chaotic behavior of the system. Analysis of level statistics 
of random quantum circuit which hosts chaotic and localized phases supports the claim that $\beta$-$h$ model 
grasps universal features of level statistics in transition between ergodic and many-body localized phases 
also for systems breaking time-reversal invariance. 
 
\end{abstract}

\maketitle

\section{Introduction}

Many-body localization (MBL) \cite{Gornyi05, Basko06} manifesting ergodicity breaking in disordered interacting quantum
many-body systems \cite{Huse14, Nandkishore15, Alet18, Abanin19} has attracted a vivid attention over the last 
decade. Important results
 include an emergent integrability of MBL phase 
due to the existence of local integrals of motion (LIOMs) \cite{Serbyn13b,Huse14,Ros15,Imbrie16,Mierzejewski18} and the
associated logarithmic growth of the bipartite entanglement entropy after a  quench from a separable state
\cite{Znidaric08, Bardarson12}. A wide regime of subdiffusive transport 
on the ergodic side of the transition was found \cite{BarLev15, Luitz16b,Mierzejewski16}.
Signatures of MBL have been observed experimentally in 1D \cite{Schreiber15, Smith16} and in 2D system \cite{Choi16}
see, however, \cite{DeRoeck17}. Recently, the very 
existence of MBL in the thermodynamic limit 
has been questioned \cite{Suntais19} opening a new debate \cite{Abanin19z,Sierant19z,Panda19}. {While the status
of MBL in the thermodynamic limit is of utmost importance for the understanding of this phenomenon from purely theoretical viewpoint,
the real systems studied in this respect are finite \cite{Kondov14, Schreiber15, Choi16, Luschen17} often reaching very modest sizes
that enable precise studies \cite{Roushan17,Lukin19,Rispoli19}. In this work we concentrate on systems of such a size.}

Spectral statistics of ergodic systems with (without) time reversal invariance
follow predictions of  Gaussian orthogonal (unitary) ensemble (GOE, GUE, respectively) of random matrices
\cite{Mehtabook,Haake} while
eigenvalues of localized systems are uncorrelated resulting in Poisson statistics (PS).
A ratio of consecutive spacings between energy levels
\begin{equation}
r^{(n)}_i = \min \{ \frac{E_{i+2n}-E_{i+n}}{E_{i+n}-E_i}, \frac{E_{i+n}-E_i}{E_{i+2n}-E_{i+n}} \}
\label{higher}
\end{equation}
was proposed  as a simple 
probe of the level statistics in \cite{Oganesyan07} with $n=1$
and employed in investigation of ergodicity breaking in various settings
\cite{Pal10, Santos10, Mondaini15,Luitz15,Luitz16,Janarek18,Wiater18,Mace19,Sierant19c}. 
Higher order spacing ratios ($n>1$), studied in
\cite{Chavda14,Harshini18,Kota18,Bhosale19,Buijsman18}, are valuable tools to assess 
properties of level statistics. In contrast to standard measures such as level spacing 
distribution or number variance  \cite{Mehtabook, Haake} they do not require the so called unfolding, i.e. the procedure
of setting the density of energy levels 
$\rho(E)$ to unity which can lead to misleading results \cite{Gomez02}.
Recently, an analytical understanding of an appearance of 
random matrix theory statistics in systems 
without a clear semiclassical limit have been developed
in a periodically driven Ising models \cite{Kos18,Bertini18}  or in random Floquet circuits \cite{Chan18a}. 
Variants of such systems have been argued to undergo ergodic-MBL transition
\cite{Chan18, Braun19}.

In this work we introduce a two-parameter $\beta$-$h$ model which assumes
a level repulsion determined by exponent $\beta$ between $h$ neighboring eigenvalues. 
{Our model is a natural
extension of the so called $\beta$-Gaussian model \cite{Buijsman18} claimed to represent the level statistics in the transition to MBL.
We show that the second parameter, the interaction distance $h$ is essential for understanding the transition and reproducing the
numerical results obtained for various physical models. In particular,} 
we demonstrate that distributions of higher order spacing ratios $r^{(n)}$ across the 
ergodic-MBL transition
in disordered XXZ spin chain are faithfully captured by $\beta$-$h$ model and the obtained 
$\beta$ and $h$ parameters provide a simple perspective on
short-range and long-range spectral correlations. 
The latter, captured effectively by the interaction range $h$,
are further investigated by means of the SFF revealing a link between $h$ 
and the Thouless time. {We demonstrate that} $\beta$-Gaussian model fails to describe long-range spectral correlations.
 An analysis of a local Haar-random unitary nearest-neighbor quantum
circuit system introduced in \cite{Chan18} indicates that also in such a generic system 
the  spectral statistics can be grasped with the $\beta$-$h$ model
demonstrating the robustness of observed features of level statistics.

{The paper is organized as follows. In Section~\ref{sec1} we introduce the  $\beta$-$h$ model and discuss properties of its level statistics. In Section~\ref{sec2} we show that the $\beta$-$h$ model accurately reproduces level statistics of disordered XXZ spin chain across the many-body localization transition. In Section~\ref{sec3} we show that the $\beta$-$h$ model grasps also level statistics of disordered Bose-Hubbard model. In Section~\ref{sec4} we demonstrate that $\beta$-$h$ applies also to ergodic-MBL transition in systems with broken time-reversal symmetry and without local conservation laws by considering level statistics of a random quantum circuit. We conclude in Section~\ref{sec5}. }

\section{The $\beta$-$h$ model} 
\label{sec1}

The joint probablity density function (JPDF) of eigenvalues of matrix from GOE (GUE) with $\beta=1$ ($\beta=2$)
can be written as a partition function of a fictitious 1D gas of particles
$\mathcal P (E_1,..., E_N) = Z_N^{-1}  \mathrm{e}^{-\beta \mathcal{E}(E_1,..., E_N)} $
where $Z_N$ is a normalization constant and the energy $\mathcal{E}$ includes a trapping potential
$U(E) \propto E^2$ and pairwise logarithmic interactions
$V(|E-E'|) = -\log(|E-E'|)$.
Eigenvalues
$E_1 < ... < E_N$
lie on a ring of length $N$
which
confines them rendering the trapping potential $U(E)$ unnecessary.
The JPDF can be written as
\begin{equation}
 \mathcal P_h^{\beta} (E_1,..., E_N) = Z_N^{-1}  \prod_{i=0}^{N} |E_i - E_{i+1}|^{\beta} ... |E_i - E_{i+h}|^{\beta}.
 \label{eq: SRPM}
\end{equation}
The GOE (GUE) case is obtained when $h\rightarrow \infty$ with the appropriate value of $\beta$.
The form of \eqref{eq: SRPM} suggests various  models of intermediate level statistics 
 between GOE (GUE) and PS. 
For instance, one can keep $h\rightarrow \infty $ and vary $\beta$, obtaining the so called $\beta$-Gaussian ensemble
\cite{Buijsman18}.
When $h$ is an integer number which sets the number of correlated neighboring eigenvalues
one arrives at the so called short-range plasma model introduced in \cite{Bogomolny01} 
(see also \cite{Pandey19a, Pandey19b}). 

In this work we extend this model by allowing $h$ to be a real number. 
Denoting by $\left \lfloor{.}\right \rfloor $ the floor function, the factor in \eqref{eq: SRPM}
becomes
$|E_i - E_{i+1}|^{\beta} ... |E_i - E_{i+\left \lfloor{h}\right \rfloor}
|^{\beta}|E_i - E_{i+\left \lfloor{h}\right \rfloor+1}|^{\beta(h-\left \lfloor{h}\right \rfloor)}$,
hence defining the $\beta$-$h$ model where $h \in[1,\infty)$ and $\beta \in [0,1]$ ($\beta \in [0,2]$) for GOE(GUE)-PS 
transition.
Varying continuously $h$ and $\beta$ allows us to capture spectral statistics of disordered 
quantum many-body systems across the ergodic-MBL transition, 
while a simple form of JDPF of $\beta$-$h$ model
yields insight into correlations between eigenvalues.
Semi-analytical results for $\beta-h$ model are available only for integer values of $h$ and $\beta$ 
\cite{Bogomolny01}. In particular,  the number variance, {defined as the variance of the number of eigenvalues in an interval $(E, E+L)$}  reads
\begin{equation}
 \Sigma^2(L) =\chi L,
 \label{sigmaSR}
\end{equation}
for $L\gg1$
where $\chi = 1/( \beta h+1)$. The spectral rigidity of GOE (GUE) which manifests
itself in the logarithmic growth of the  variance $\Sigma^2(L)$
is replaced by a finite spectral compressibility $\chi$.
Thus, a profound change in long-range spectral correlations happens when $h<\infty$.
Interestingly, we find that \eqref{sigmaSR} is fulfilled with an excellent agreement
for $\beta$-$h$ model as our Monte Carlo simulations (obtained sampling JPDF of $\beta$-$h$ model with the 
Metropolis-Hastings algorithm \cite{Hastings70})
shows for arbitrary real $\beta\in[0,2]$ and $h\in[1,40]$. 

\begin{figure}
 \includegraphics[width=0.9\linewidth]{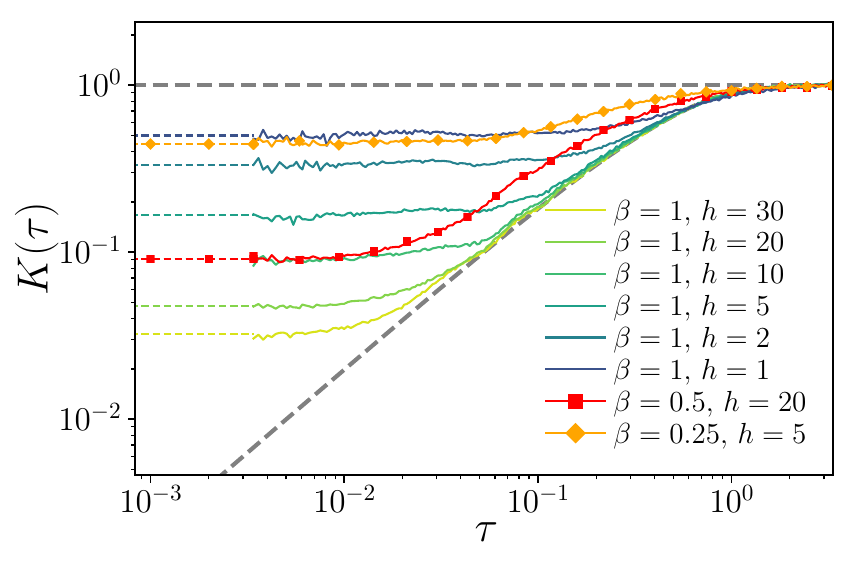}
  \caption{ {  The SFF  $K(\tau)$ of $\beta$-$h$ model.
  For $\tau <0.003$, $K(\tau)$ was replaced by analytically determined value of
  $K(0)$. Grey dashed lines correspond to GOE and PS.
 }} \label{sffBH}
\end{figure}

%A number of analytical results is available for $h=1$ \cite{Atas13, Atas13b}. 
A straightforward 
application
of the method of \cite{Bogomolny01} shows that distributions of higher order spacing ratios $P(r^{(n)})$ for $h=1$
are given by 
\begin{equation}
 P(r^{(n)}) =  \mathcal{N}_{n,\beta}\frac{ (r^{(n)})^{ \beta + (n-1) (\beta+1) }  }{ (1+r^{(n)})^{2(\beta+1)n} }
\end{equation}
where $\mathcal{N}_{n,\beta}= [_{2}F_{1}(n(1+\beta),2n(1+\beta),1+n(\beta+1),-1)/(\beta+1)n ]^{-1}$
is a normalization constant and $_{2}F_{1}$ is Gauss hypergeometric function.
 Such distributions of higher order spacing ratios $P(r^{(n)})$ at $h=1$ constitute a very good approximation 
 for systems close to the MBL phase where $h\approx 1$ and 
provides  analytical expressions for average higher order spacing ratios $\overline r^{(n)}$ (including PS for $\beta=0$). {To obtain $P(r^{(n)})$ for arbitrary $h \in[1,\infty)$ and $\beta \in [0,1]$ we again
sample JPDF of $\beta$-$h$ model with the Monte Carlo approach.}

\subsection{Spectral form factor of $\beta$-$h$ model}

Consider the \textit{spectral form factor} (SFF) \cite{Mehtabook, Haake}:
\begin{equation}
 K(\tau)  = \frac{1}{Z}\left \langle \left| \sum_j g( \epsilon_j) \mathrm{e}^{-i E_j \tau} \right|^2 \right \rangle,
 \label{eq: Kt}
\end{equation}
where $Z$ assures that $K(\tau) \stackrel{\tau \rightarrow \infty}{\rightarrow} 1$,
the spectrum is unfolded (for remarks on unfolding see Appendix~\ref{appen})  and $g(\epsilon)$ is a Gaussian function which 
vanishes at the edges of spectrum reducing their influence  -- see also Appendix.
The SFF allows to identify two important time scales in disordered systems:
the Heisenberg time $\tau_H$
equal to the inverse level spacing, 
 beyond which the discrete nature of the energy spectrum manifests itself  and the Thouless time $\tau_{Th}$, which is 
the time scale beyond
which SFF admits universal GOE (GUE) form $K(\tau)\approx 2\tau$ \cite{Chan18, Suntais19}.
The existence of  two time scales is %immediately reminiscent of 
reflected in 
the JPDF of 
$\beta$-$h$ model, where the correlations between eigenvalues are of the GOE (GUE) form
on energy scales smaller than $h$ level spacings so that
$\tau_{Th}$, inversely proportional to $h$, (for $\beta=1,2$)
provides
 a physical interpretation of the interaction range $h$ in $\beta$-$h$ model. 

\begin{figure}
 \includegraphics[width=0.9\linewidth]{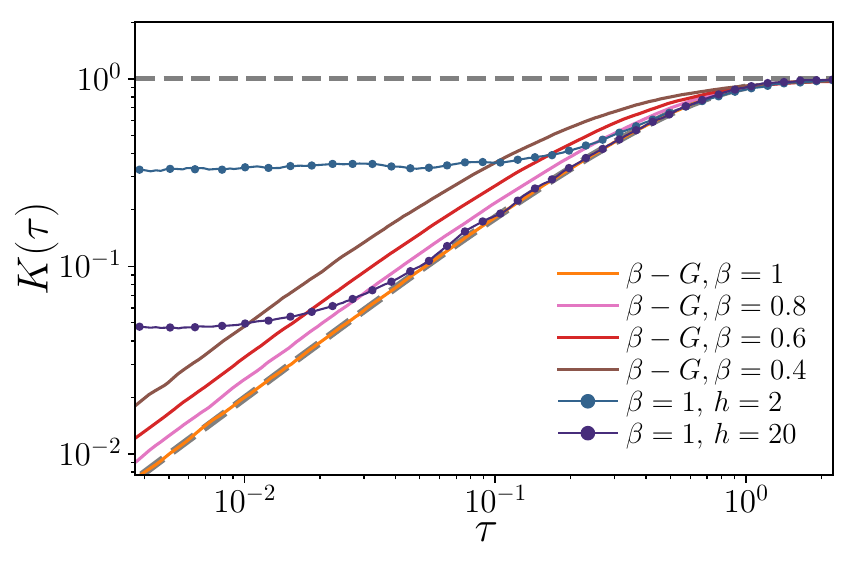}
  \caption{ {  Comparison of SFFs of $\beta-h$ model (lines with dots) and of $\beta$-Gaussian ensemble (solid lines).
Grey dashed lines correspond to GOE and PS.
 }} \label{sffBH2}
\end{figure}
SFF of $\beta$-$h$ model is shown in Fig.~\ref{sffBH}.
 For $\beta=1$, SFF of $\beta$-$h$ model follows prediction for GOE down to Thouless time $\tau_{Th}$ which
depends on the interaction range $h$ (roughly: $\tau_{Th}\approx 2/(h+1)$). SFF for $\beta<1$ shows that
it is possible to have spectral statistics with $h>1$ and $\tau_{Th}=\tau_H=1$. 

We note that \eqref{sigmaSR} implies that $K(0) = 1/(\beta h +1)$ -- an analytical prediction for integer $\beta$ and $h$
which is very well confirmed by numerical data for arbitrary $\beta$ and $h$ as shown in Fig.~\ref{sffBH}.

Fig.~\ref{sffBH2} shows comparison of SFFs of $\beta-h$ model and of $\beta$-Gaussian ensemble. The two parameters
of the $\beta-h$ model allow to reproduce the typical behavior of SFF of a disordered many-body
system across ergodic-MBL transition { \cite{Suntais19, Sierant19z} }-- the Thouless time $t_{Th}$ {of a physical quantum many-body system increases with disorder strength $W$, which, in $\beta$-$h$ model} is {reflected} by a
decreasing range $h$ of eigenvalue interactions. The vanishing level repulsion in the vicinity of the localized 
phase is reflected by sufficiently small value of the exponent $\beta$. This is not the case for 
the $\beta$-Gaussian ensemble. As soon as $\beta<1$, the deviation from the SFF of GOE is observed for all 
$\tau <1$ as Fig.~\ref{sffBH2} illustrates. Thus, the $\beta$-Gaussian ensemble is unable of reproducing 
the typical behavior of SFF, $K(\tau)$, in disordered system in which $K(\tau)$
deviates from the universal GOE curve only for times smaller than the Thouless time 
$t_{Th}$. Moreover, for $\beta<1$ the predictions of $\beta$-Gaussian ensemble fail to reproduce a small $\tau$
behaviour showing a rapid decrease with decreasing $\tau$ instead of a saturation as expected closer to Poisson regime.

\section{ XXZ spin chain} 
\label{sec2}

Le us go beyond a comparison of the statistical models among themselves and compare their predictions with different physical models.
As a starting point for testing purposes we consider a standard disordered XXZ spin-1/2 chain with Hamiltonian
given by
\begin{equation}
 H= J\sum_{i=1}^{L} \ \vec{S}_i \cdot \vec{S}_{i+1} + \sum_{i=1}^{L} h_i S^z_i,
 \label{eq: XXZ}
\end{equation}
where  $\vec{S}_i$ are spin-1/2 matrices, $J=1$ is fixed as the energy unit,
periodic boundary conditions are assumed $\vec{S}_{L+1} = \vec{S}_1$ and $h_i \in [-W, W]$ 
are independent, uniformly distributed random variables.
The model \eqref{eq: XXZ} has been widely studied 
in the MBL context
\cite{Znidaric08, Pal10, Berkelbach10,Luitz15, Agarwal15, Bera15, Enss17, Bera17, Herviou19, Colmenarez19},
and its level statistics have been addressed in
\cite{Serbyn16,Bertrand16,Kjall18, Sierant19b}.
Recently, the $\beta$-Gaussian ensemble was suggested to describe  the ergodic-MBL transition \cite{Buijsman18}. 
As {the analysis of Section~\ref{sec1} has shown} this claim is questionable. Further, we shall show 
that {the  $\beta$-Gaussian ensemble}  reproduces level correlations only 
on a single level spacing scale while missing
longer-range spectral correlations. 
Both aspects of level statistics are grasped by $\beta-h$ model.

\begin{figure}
 \includegraphics[width=1.0\linewidth]{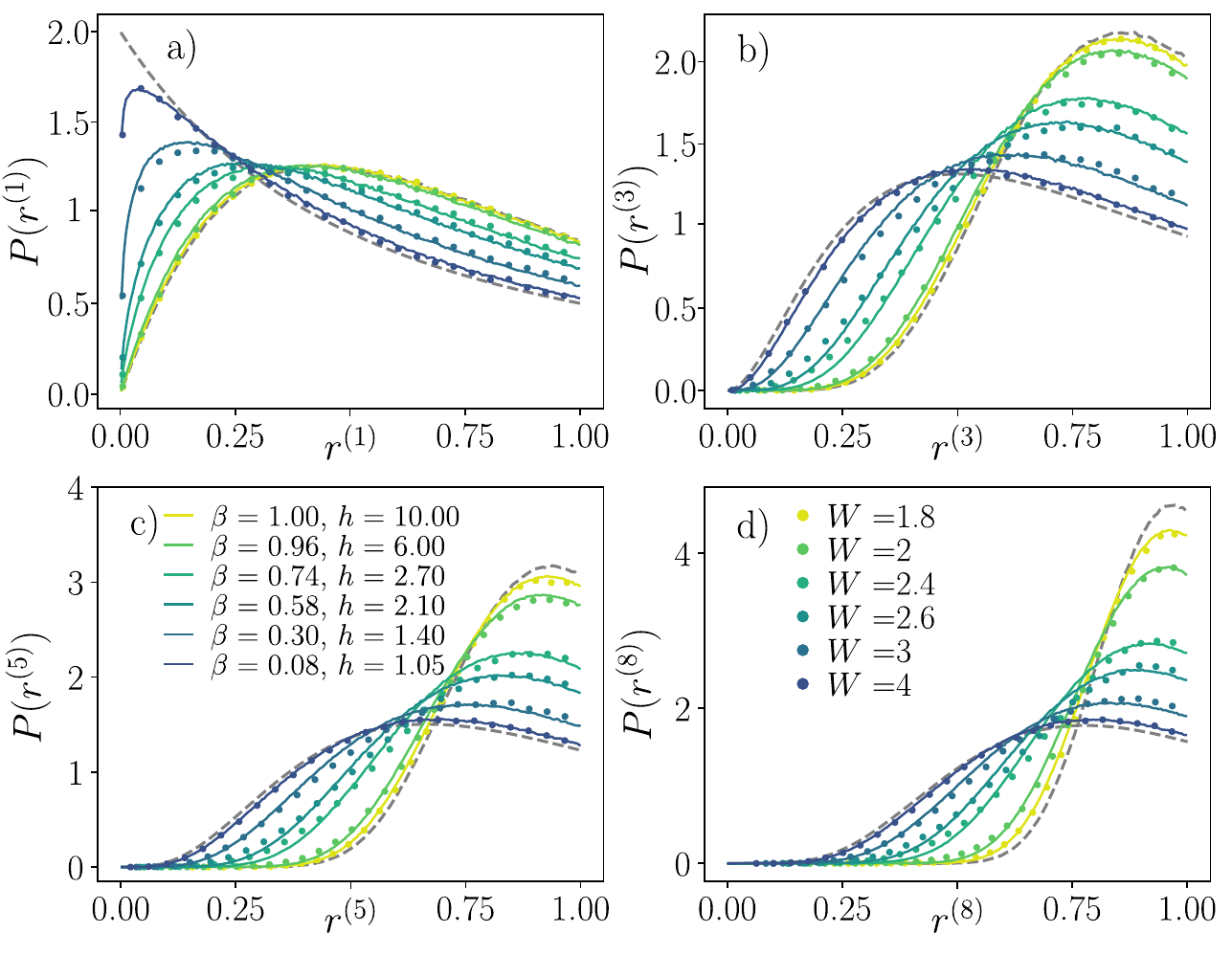}
  \vspace{-0.65cm}
  \caption{ { Distributions of higher order spacing ratios of disordered XXZ spin
  chain \eqref{eq: XXZ} of size $L=18$ for various disorder strengths $W$ are denoted by symbols.
  Lines correspond to $\beta$-$h$ model with parameters shown in panel c). Grey dashed lines correspond
  to $P(r^{(n)})$ distributions for GOE and PS.
  \label{fig: higher_spacings} 
 }}
\end{figure}
\begin{figure}
 \includegraphics[width=0.99\linewidth]{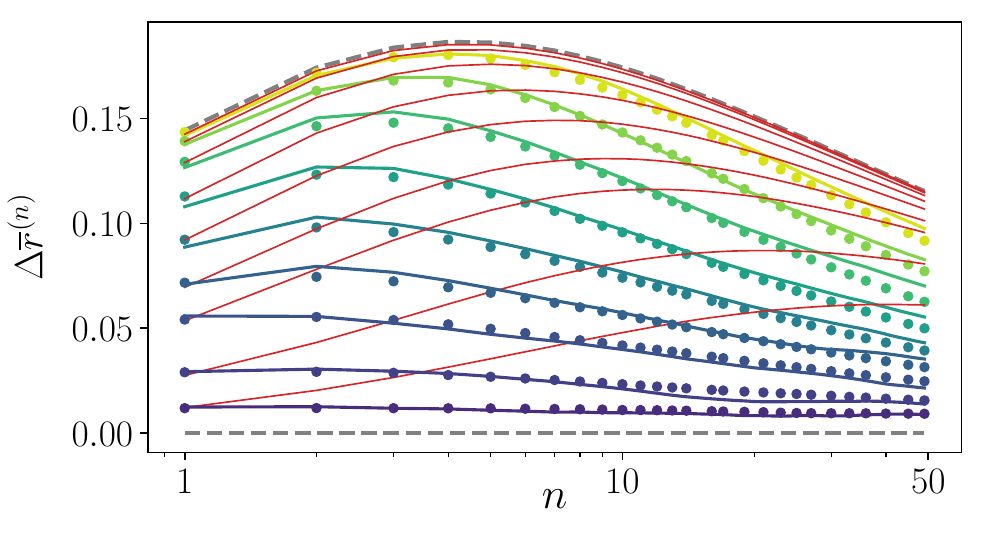}
 \vspace{-0.45cm}
  \caption{ { Symbols show the average higher order gap ratios $\Delta \overline r^{(n)}$ (see text) as function of
  $n$ for disorder strengths $W=1.8, 2, 2.2, 2.4, 2.6, 2.8, 3, 3.4,4$ (from top to bottom)  for XXZ chain of size $L=18$.
  Corresponding fits of $\beta$-$h$ model are drawn
  by solid lines, the $\beta$, $h$ parameters are the same as in Fig.~\ref{fig: higher_spacings}, additional
  $W=2.2,2.8, 3.4$, are fitted by $\beta=0.90, 0.46, 0.18$ and $h=3.60, 1.70, 1.30$ respectively.
  Red lines correspond to $\beta$-Gaussian ensemble with $\beta=0.98, 0.94, 0.84, 0.68, 0.52, 0.38, 0.26, 0.12, 0.05$
  (from top to bottom).
  Grey dashed lines: $\Delta \overline r^{(n)}$ for GOE and PS respectively.
 }}\label{fig: avrXXZ} 
\end{figure}

Eigenvalues of the XXZ spin chain \eqref{eq: XXZ} are obtained by an exact diagonalization 
for small sizes or with shift-and-invert method \cite{Pietracaprina18}
 for 
$L=18,20$. For each $W$ we accumulate eigenvalues from 
$2000$ ($400$) disorder realizations for $L \leq 18$ ($L=20$). 
The  higher order spacing ratios \eqref{higher} are calculated 
using $500$ eigenvalues from the middle of the spectrum. The resulting 
exemplary distributions $P(r^{(n)})$ of higher order spacing  ratios
{for $n=1, 3, 5, 8$} 
are shown in Fig.~\ref{fig: higher_spacings}. Parameters for $\beta$-$h$
model are obtained by minimizing the deviation between $P(r^{(n)})$ distributions for XXZ spin chain
and $\beta$-$h$ model. 
A very good agreement between the 
distributions obtained for the model \eqref{eq: XXZ} and predictions of $\beta$-$h$ model is observed
in the whole transition region between the ergodic and MBL phases. Note that both parameters $\beta$ and $h$ are needed to
reproduce $P(r^{(n)})$ distributions for $n\geq 1$.
To demonstrate that the agreement between $\beta$-$h$ model and level statistics of
XXZ spin chain in ergodic-MBL crossover persists to larger energy scales,
 we calculate 
$\Delta \overline r^{(n)}=\overline r^{(n)}-\overline r^{(n)}_{PS}$,
where $\overline r^{(n)}$ is the average value of $n$'th order spacing ratio $\overline r^{(n)}$ and
$\overline r^{(n)}_{PS}$ is the $n$'th order average gap ratio for PS.
The resulting values of $\Delta \overline r^{(n)}$ as function of $n$ are shown in Fig.~\ref{fig: avrXXZ}.
Even though the parameters of $\beta$-$h$ model are determined by fit of
$P(r^{(n)})$ for $n=1, 3, 5, 8$ only  (fits for all $n$ taken with the same weight), 
the good agreement between $\Delta \overline r^{(n)}$ for 
XXZ spin chain and for $\beta$-$h$ model persists up to $n=50$. 
Interestingly,
on the ergodic side of the transition, for $W \leq 2.4$, the values of $\Delta \overline r^{(n)}$
for $n\geq 20$ predicted by $\beta$-$h$ model are consequently overestimating the values for XXZ spin chain. 
Since the larger value of $\Delta \overline r^{(n)}$ implies stronger level correlations at the scale determined by $n$,
this means that energy levels of $\beta$-$h$ model, not coupled directly in JPDF \eqref{eq: XXZ}, are still correlated
more strongly than energy levels of the system across the ergodic-MBL transition.
For comparison, we {also} show {the} predictions of $\beta$-Gaussian ensemble \cite{Buijsman18} in Fig.~\ref{fig: avrXXZ}.
Only the values of $\Delta \overline r^{(1)}$ are well reproduced by this approach, 
for $n\geq2$), the values of  $\Delta \overline r^{(n)}$ are %severely 
overestimated showing 
that finite $h$ is an essential feature of level statistics in the MBL transition.

\subsection{Scaling of level repulsion exponent $\beta$ and range of interactions $h$ at the ergodic-MBL transition}

\begin{figure}
\includegraphics[width=1\linewidth]{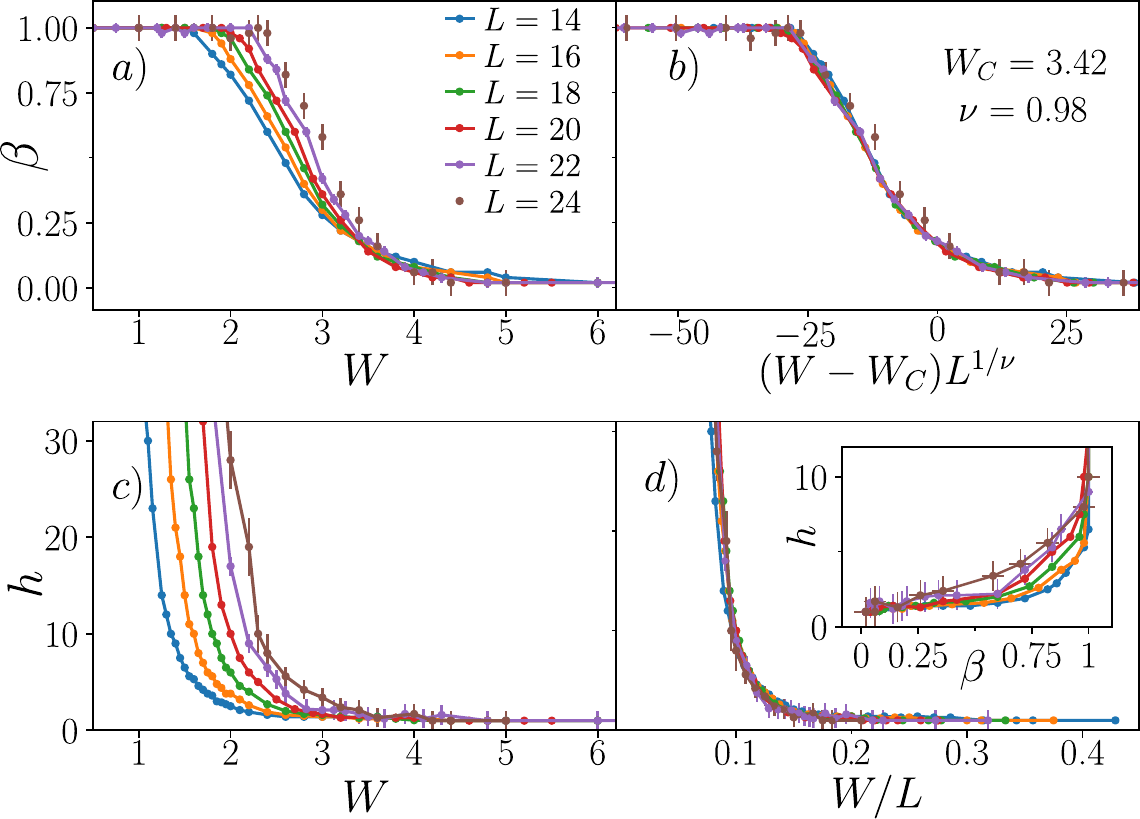}\vspace{-0.4cm}
  \caption{  Left: level repulsion exponent 
  $\beta$ -- a) and the range $h$ of interactions of eigenvalues  -- c)
  as a function of disorder strength $W$ in XXZ chain. Right: The collapse of the
  data for $\beta(W)$ upon rescaling $W\rightarrow(W-W_C)L^{1/\nu}$ -- b) and for $h(W)$ using 
 $W\rightarrow W/L$ rescaling.
  Inset in panel d) shows the 
  dependence $h(\beta)$ for various system sizes.
 } \label{fig: bhXXZ} 
\end{figure}

The  $\beta$ and $h$ parameters characterizing
level statistics across the ergodic-MBL transition
are shown in Fig.~\ref{fig: bhXXZ}.
In the ergodic phase, at small disorder strengths $W$, GOE describes level statistics well, 
hence $\beta=1$ and $h\rightarrow\infty$. 
Upon increase of $W$, the range of 
interactions $h$ and the level repulsion exponent $\beta$ decrease leading to PS 
for sufficiently strong disorder.
Notably, the system size dependences 
of $h(W)$ and $\beta(W)$ are very different. 
The data for $\beta(W)$ collapse upon rescaling $W\rightarrow(W-W_C)L^{1/\nu}$ with 
$W_C\approx3.4$ and $\nu \approx 1$,  
similarly
to %the scaling form used in \cite{Luitz15} for the average gap ratio $\overline r^{(1)}$ 
the average gap ratio $\overline r^{(1)}$ \cite{Luitz15} 
indicating that $\overline r^{(1)}(W)$ and $\beta(W)$ contain similar information. In particular, 
both measures lead to the exponent $\nu < 2$ violating the Harris bound
\cite{Harris74, Chayes86, Chandran15a}.
On the other hand,
data for $h(W)$ collapse upon rescaling $W\rightarrow W/L$.
As inset in Fig.~\ref{fig: bhXXZ} d) demonstrates, the decrease of the 
level repulsion exponent $\beta$ in the transition region 
is accompanied by the interaction range $h$ increasing with $L$ for a given value
of $\beta$. 
Therefore, our data indicate the presence of the transition to MBL phase with vanishing 
level repulsion exponent $\beta=0$ at disorder strength 
$W_C$ even though 
the interaction range admits a certain fixed value $h_0=h(W^*)$
at disorder strength $W^*$ increasing linearly with the system size $L$.
The linear dependence $W^* \sim L$ 
persists steadily up to the largest available system size $L=24$ but since the values 
of $h$ in the transition region do not exceed $10$ we cannot
conclude whether $h$ diverges or stays finite at the transition in the thermodynamic limit.
Nevertheless, in either case, the level repulsion vanishes at the transition
in $L\rightarrow \infty$ limit, in accordance with recent phenomenological treatments
\cite{Thiery18, Morningstar19}.
We note that 
a linear with system size dependence 
for deviation of 
$\overline r^{(1)}$ from the value characteristic for GOE was found recently 
in \cite{Suntais19}. This observation is related to our finding that 
$W^* \sim L$ since disorder strength for which $h$ becomes of the order of e. g. $h_0=10$
is, at the same time, the moment for which $\overline r^{(1)}$ departs from 
the value characteristic for GOE. 
While $\beta-h$ model puts the long-range spectral statistics examined in 
\cite{Suntais19} in another perspective, we emphasize that the presented
data suggest transition to MBL phase at disorder strength $W_C$ in the thermodynamic limit.

\subsection{The spectral form factor XXZ spin chain}

\begin{figure}
 \includegraphics[width=0.95\linewidth]{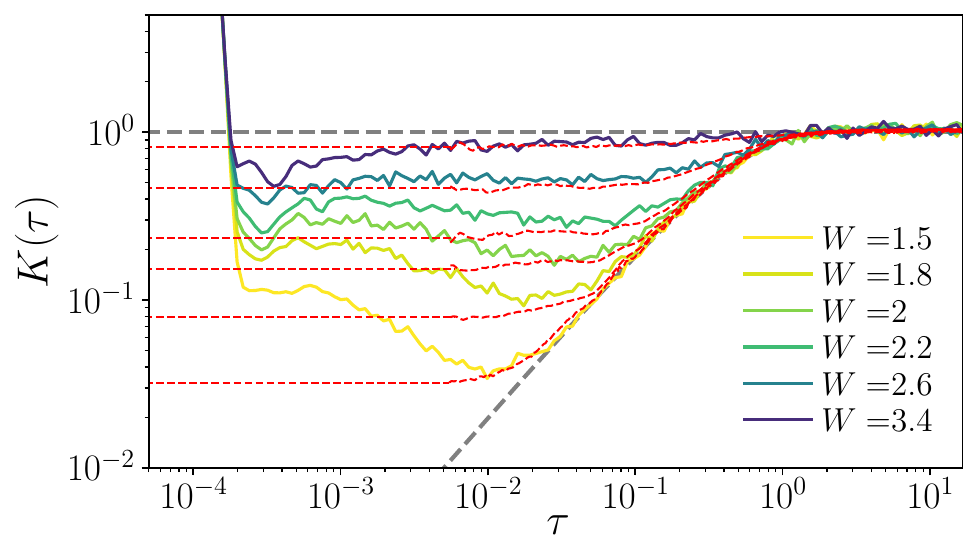}
 \vspace{-0.6cm}
  \caption{  SFF for system \eqref{fig: bhXXZ} of size $L=18$ 
  for various disorder strengths $W$. Predictions of $\beta$-$h$ model with parameters the same as in Fig.~\ref{fig: avrXXZ}
  (data for $W=1.5$ fitted with $\beta=1$, $h=30$) are denoted by red 
  dashed lines (for $\tau < 0.005$ SFF was replaced by exact value in $\tau=0$).
  Grey dashed lines correspond to GOE and PS.
 } \label{fig: ktXXZ}
\end{figure}
Let us now consider the SFF
 of XXZ spin chain {which} is shown in Fig.~\ref{fig: ktXXZ} along with {the} predictions of $\beta$-$h$
model. Beyond the Heisenberg time $\tau_H$, $K(\tau)=1$. For smaller $\tau$, the SFF of XXZ spin chain follows the GOE prediction 
down to the Thouless time $\tau_{Th}$ which {increases} monotonically with disorder strength $W$. 
The behavior is captured by SFF of the $\beta$-$h$ model. For $ \tau < \tau_{Th}$, an increase in the SFF of XXZ chain 
is observed for disorder strengths $W$ corresponding to the ergodic side of the transition,
whereas SFF remains constant for the $\beta$-$h$ model. 
The latter behavior signals weak correlations between eigenvalues of $\beta$-$h$ model beyond energy scale determined
by $h$, whereas the behavior of SFF of XXZ spin chain indicates even weaker correlations of its eigenvalues.

\subsection{Level statistics and number variance across the ergodic-MBL transition}

\begin{figure}
 \includegraphics[width=0.9\linewidth]{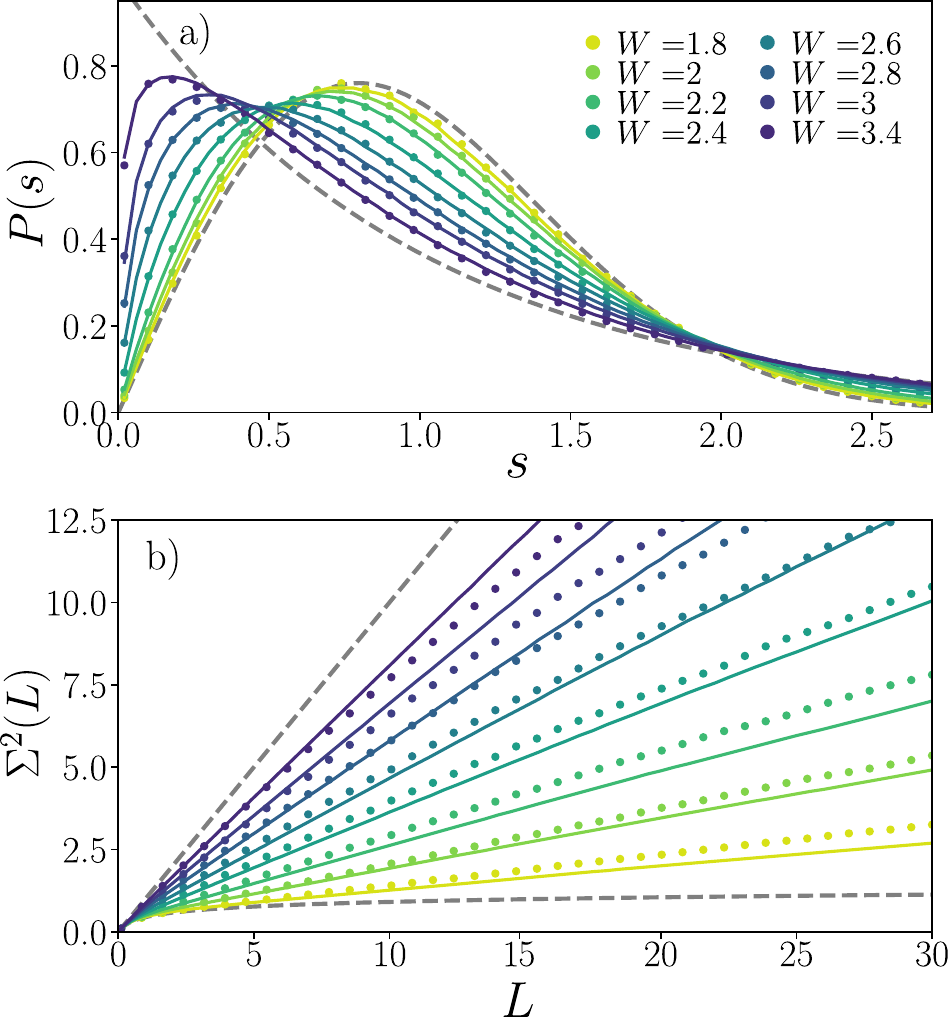}
  \caption{ { Panel a) -- level spacing distribution $P(s)$ of XXZ spin chain 
  of size $L=18$ for various disorder strengths $W$ are denoted by markers.
  Lines correspond to predictions of $\beta$-$h$ model with parameters as in the main text.
  Panel b) -- number variance $\Sigma^2(L)$ of XXZ spin chain and prediction of $\beta$-$h$ model.
 }} \label{figlsnv}
\end{figure}

We revisit now the
level spacing distribution and the number variance of the disordered XXZ spin chain 
in the ergodic-MBL crossover,  shown in Fig.~\ref{figlsnv}.
Level spacing distributions are very faithfully reproduced by
$\beta$-$h$ model in the whole crossover regime. There are, however slight deviations
in the number variance $\Sigma^2(L)$  of XXZ spin chain and $\beta$-$h$ model. 
On the ergodic side of the crossover ($W<2.4$) the $\beta$-$h$ model underestimates 
number variance of XXZ spin chain indicating weaker long-range spectral correlations of the 
latter, in agreement with the analysis of $\Delta \overline r^{(n)}$ in this regime. 
For large $W$, the prediction of $\beta$-$h$ model overestimates the number variance
of XXZ spin chain --that is probably related to effects of a finite number of eigenvalues 
$n_e$ from a single disorder realization which are known to contribute as $-L^2/n_e$ to
the number variance $\Sigma^2(L)$ \cite{Sierant19b}.

We note  that results for the number variance $\Sigma^2(L)$ are strongly dependent on the way the 
level unfolding is performed, in view of that we conclude that the higher order spacing ratios are
more reliable in extracting informations about spectral correlations beyond single level spacing.
Furthermore, when one inspects tails $s\gtrsim4$ of the level spacing distribution on the logarithmic scale, 
deviations from predictions of $\beta$-$h$ model are found. As it was demonstrated in \cite{Sierant19b},
such a behavior at large $s$ is associated with the large inter-sample randomness associated with ergodic-MBL 
transition in random potentials.

 The weighted short-range plasma model for level statistics in ergodic-MBL crossover
was considered in \cite{Sierant19b}.
This  model 
takes into account inter-sample randomness, an important 
feature of MBL transition in random potentials \cite{Khemani17a}.
The JPDF of weighted short-range plasma model
is a weighted superposition of JPDF of the form \eqref{eq: SRPM} and as such
it is related to $\beta$-$h$ model. However, the necessity of reproducing the inter-sample randomness
requires an introduction of many weight parameters. This makes
use of the  weighted short-range plasma model complicated.  The simple picture of changes of interaction 
range between eigenvalues and its relation to Thouless time cannot be easily extracted
due to the complexity of the model. It must be noted however, that taking into account the
inter-sample randomness determined by a sample-averaged spacing ratio 
\cite{Sierant19b} could diminish the (small) deviations in $P(r^{(n)})$ between $\beta$-$h$ model
and XXZ spin chain for disorder strengths $W=2.4,2.6$ for which the inter-sample randomness is the largest at $L=18$.

A two-stage \cite{Serbyn16} picture of flow {of level statistics between GOE and PS} proposes that on the ergodic side of the crossover level statistics
are described by {a} plasma model with power-law interactions between eigenvalues which yields the 
following expressions for the level spacing distribution and the number variance:
\begin{equation}
P(s) = C_1 s^{\beta} \mathrm{e}^{-C_2 s^{2-\gamma}} \,\,\,\,\, \mathrm{and} \,\,\,\,\,\, \Sigma_2(L) \propto L^{\gamma}
 \label{eq: SM1}
\end{equation}
with $C_{1,2}$ determined by normalization conditions $\langle 1 \rangle=\langle s \rangle = 1$.
The exponent $\beta$ and $\gamma$ play a role similar to $\beta$ and $h$ parameters of the $\beta$-$h$ model.
In the transition from extended to localized regime in the first stage $\gamma$ changes from 0 
to 1 leading to Poissonian tail 
of $P(s)$ followed at the second stage by a change of level repulsion $\beta$. 
However, as demonstrated in \cite{Bertrand16}, the predictions of \eqref{eq: SM1} are not valid as the number
variance $\Sigma^2(L)$ in ergodic-MBL transition grows linearly (or superlinearly) -- see Fig.~\ref{figlsnv},
contrary to prediction of \eqref{eq: SM1} where $0<\gamma<1$ in the crossover regime.
Moreover, \eqref{eq: SM1} is obtained on the mean-field level \cite{Kravtsov94}, no other predictions
for this model such as JPDF are available. 
The second stage of the flow \cite{Serbyn16} coincides with $\beta$-$h$ model with $h=1$. However, as shown above,
the $h(\beta)$ dependence is such that the interaction range $h$ for fixed {value of} $\beta$ 
{is} increasing, hence $h$ becomes equal to unity only deep in the  MBL regime.

\section{Level statistics of disordered Bose-Hubbard model}
\label{sec3}

To provide further evidence that $\beta$-$h$ model is able to reproduce level statistics
of interacting disordered quantum many-body systems, we analyze higher order spacing ratios
in ergodic-MBL transition in a disordered Bose-Hubbard model \cite{Sierant17, Sierant18}
with Hamiltonian:
\begin{equation}
  H_{B} = -J \sum_{\langle i,j \rangle} \hat{a}^{\dag}_i\hat{a}_j +
 \frac{ U }{2} \sum_i \hat{n}_i (\hat{n}_i - 1) +
  \sum_i \mu_i \hat{n}_i,
 \label{eq: ham_BH}
\end{equation}
where  $a^{\dag}_i, a_i$ are bosonic creation and annihilation operators respectively, 
the tunneling amplitude $J=1$ sets the energy scale, $U=1$ is interaction strength and the chemical 
potential $\mu_i$ is distributed uniformly in an interval $[-W;W]$. This model undergoes 
transition to MBL phase beyond critical disorder strength $W_C$ which depends on interaction
strength $U$.

\begin{figure}
 \includegraphics[width=1.0\linewidth]{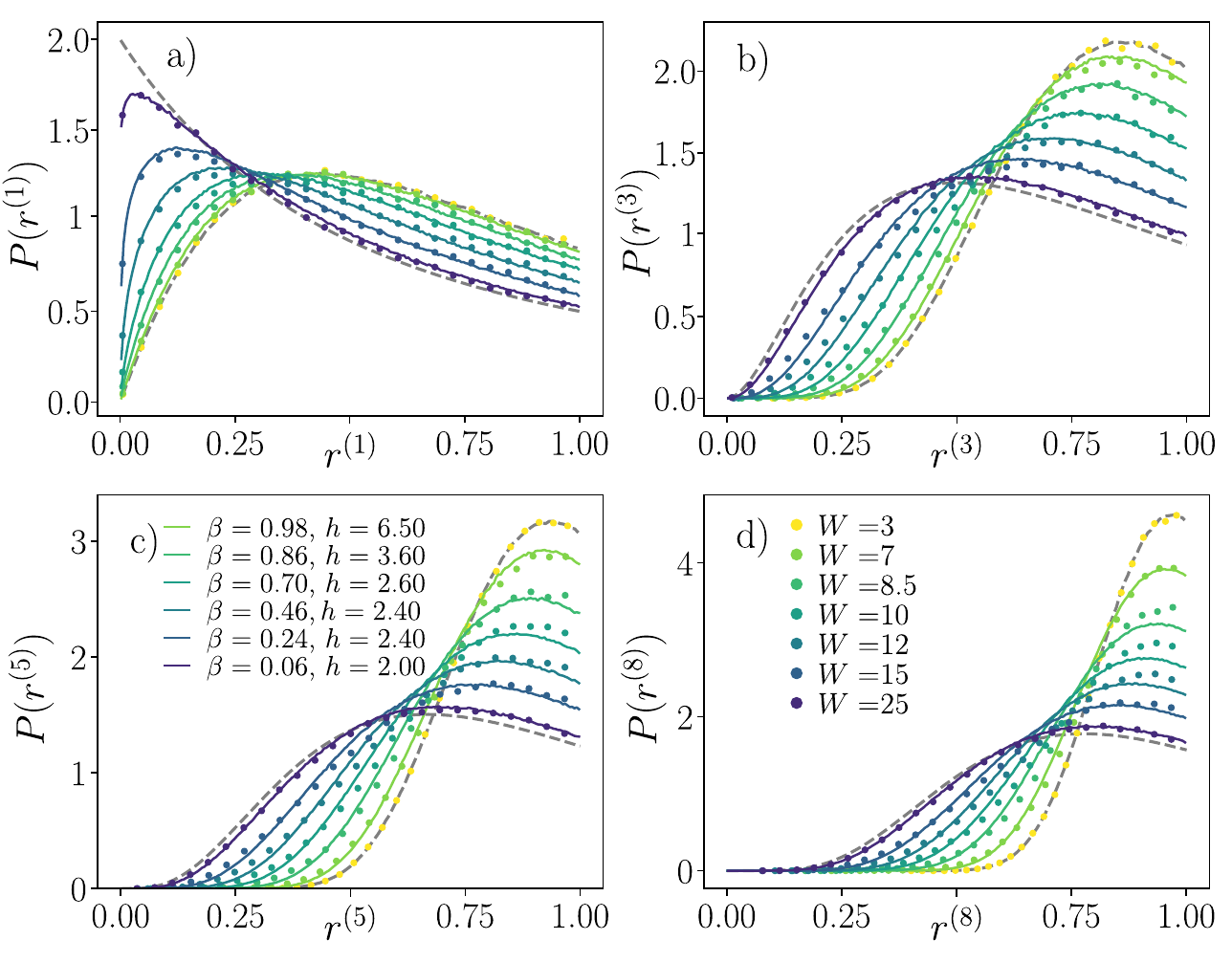}
  \caption{ { Distributions of higher order spacing ratios for Bose-Hubbard model 
  \eqref{eq: ham_BH} of size $L=8$ with $N=12$ particles are denoted by markers, fits of 
  $\beta$-$h$ model  to data with $W\geq 7$  are denoted by solid lines. Higher order spacings distributions ($n=1,3,5,8$) for $W=3$ are indistinguishable
  from appropriate distributions for GOE. Dashed lines correspond to GOE and PS.
 }} \label{higherBH}
\end{figure}
\begin{figure}
 \includegraphics[width=0.85\linewidth]{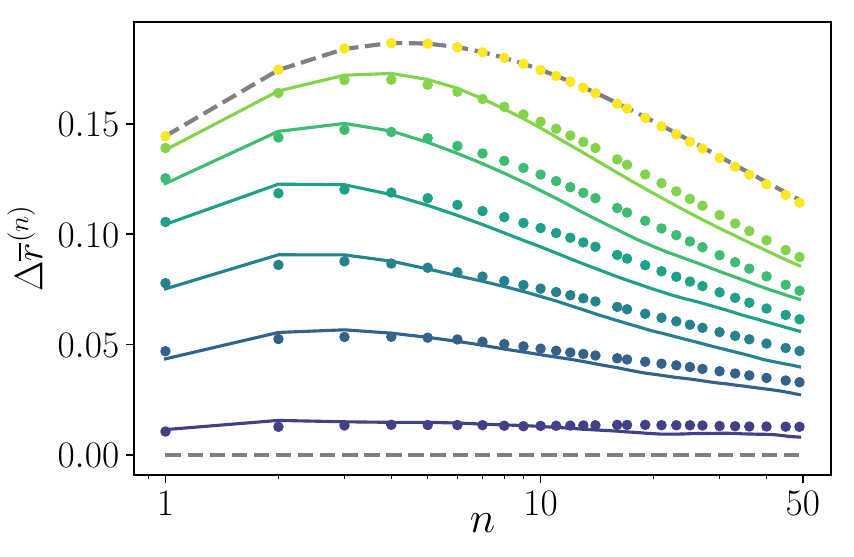}
  \caption{  Average higher order spacing ratios for Bose-Hubbard model with disorder strengths $W=3, 7, 8.5, 10, 12, 15, 25$ (from top to bottom)
  denoted
  by markers, predictions of $\beta$-$h$ model fitted to data with $W\geq 7$ with parameters the same as in Fig.~\ref{higherBH} are denoted by solid lines.
  Level statistics for $W=3$ are indistinguishable from GOE statistics on the considered energy scale.
  Grey dashed lines correspond to GOE and PS.
 }\label{higherBHavr}
\end{figure}
Distribution of higher order spacing ratios $(n=1,3,5,8)$ for the disordered 
Bose-Hubbard model are shown in Fig.~\ref{higherBH}. The $\beta$-$h$ model 
reproduces faithfully distributions $P(r^{(n)})$ in the whole crossover regime. We note that
the dependence $h(\beta)$ is markedly different as compared to the XXZ spin chain -- 
here we find $h=2$ even when the level repulsion exponent $\beta$ is close to $0$.
Average higher order spacing ratios shown in Fig.~\ref{higherBHavr} indicate that long-range 
spectral statistics are also well reproduced by the $\beta$-$h$ model. In particular, the tendency
of $\beta$-$h$ model to  overestimate long-range spectral correlations in XXZ spin chain 
is reversed in the case of Bose-Hubbard model, indicating that this is a model dependent feature.

\section{Random quantum circuit} 
\label{sec4}

Consider 1D chain of $q$-level systems of length $L$ with Floquet operator
given by \cite{Friedman19}
\begin{equation}
 W_{a_1,\ldots,a_L;a'_1,\ldots,a'_L} =U^{(1)}_{a_1, a'_1}\ldots U^{(L)}_{a_L, a'_L} \mathrm{e}^{i\sum_n \varphi_{a'_n, a'_{n+1}} },
 \label{eq: Flo}
\end{equation}
where 
$U^{(j)}$  are unitary matrices that generate 
rotations at each site, chosen independently from Haar distribution, 
$\varphi_{a_n, a_{n+1} }$ are independent Gaussian random variables with zero mean and standard deviation 
$\epsilon$ that determine coupling between neighboring sites. 
The SFF is related to the Floquet operator via 
$
K( t )=
\left \langle \mathrm{Tr}[W^{t}]\mathrm{Tr}[(W^{\dag})^{t}]   \right \rangle 
\label{eq: KtFlo}
$
where
$t$ is an integer and \eqref{eq: Kt} is recovered with $g(\epsilon)=1$ for $\tau \propto t$.
Analytic calculation \cite{Friedman19} in the limit
$q\rightarrow \infty$ shows that the system is chaotic in the thermodynamic limit  
{and}
SFF follows {the} prediction for GUE:
$K(\tau) = 2\tau$. For $q=3$, numerical calculations indicate that 
the system undergoes {a} transition between ergodic phase at $\epsilon \gtrsim 0.25$
where the statistics of eigenphases $\theta_j$ are {well described by} GUE and 
MBL phase at $\epsilon \lesssim 0.25$ with PS statistics. We now turn to 
analysis of level statistics of \eqref{eq: Flo} at finite $L$ and $q=3$.

\begin{figure}
 \includegraphics[width=1\linewidth]{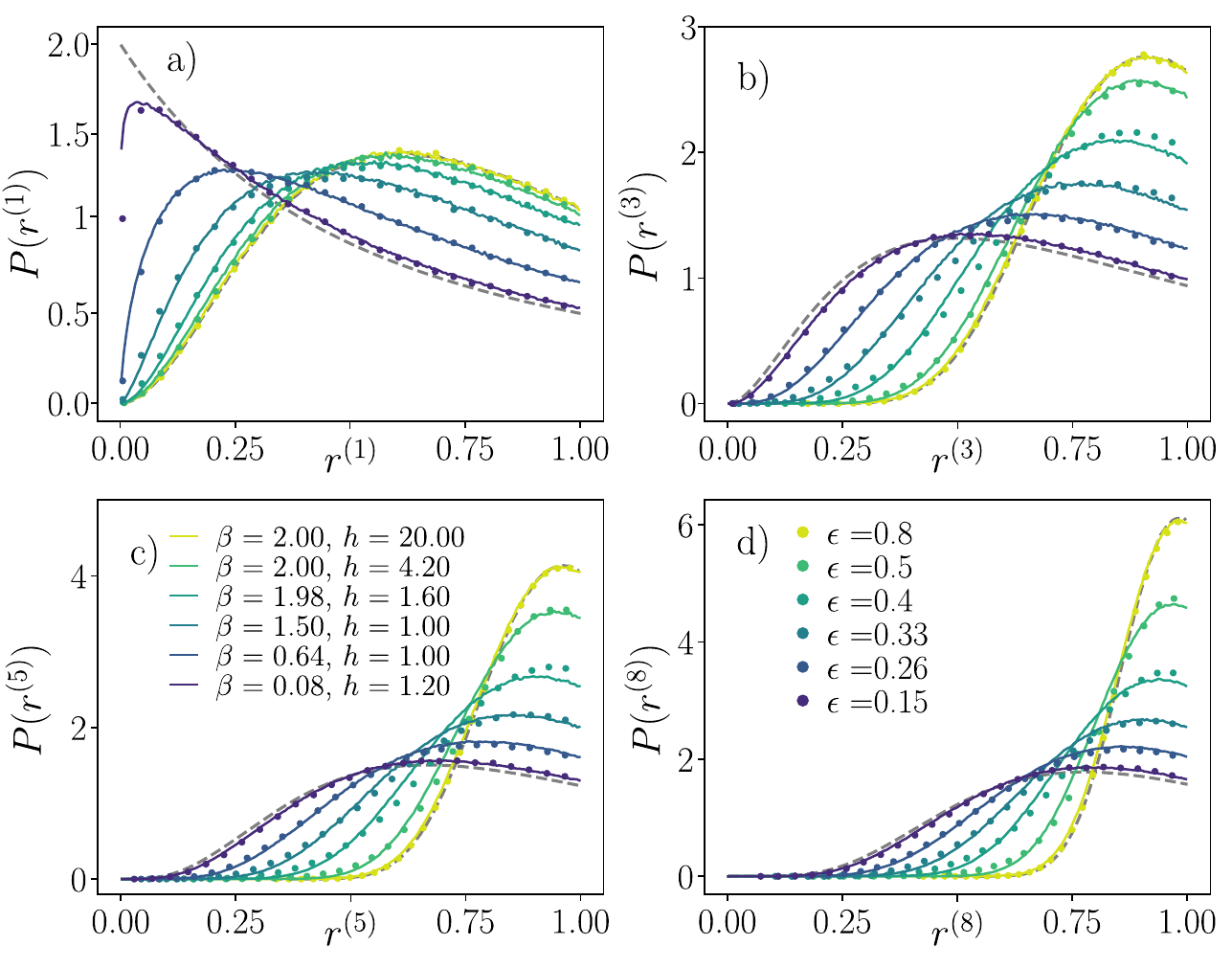}\vspace{-0.1cm}
  \caption{ { Distributions of higher order spacing ratios for model \eqref{eq: Flo} 
  with $L=8$ and $q=3$ for various $\epsilon$ are denoted by symbols.
  Lines correspond to $\beta$-$h$ model. Grey dashed lines correspond
  to $P(r^{(n)})$ distributions for GUE and PS.
  } }
  \label{fig: higher_spacingsFlo} 
\end{figure}

\begin{figure}
 \includegraphics[width=0.9\linewidth]{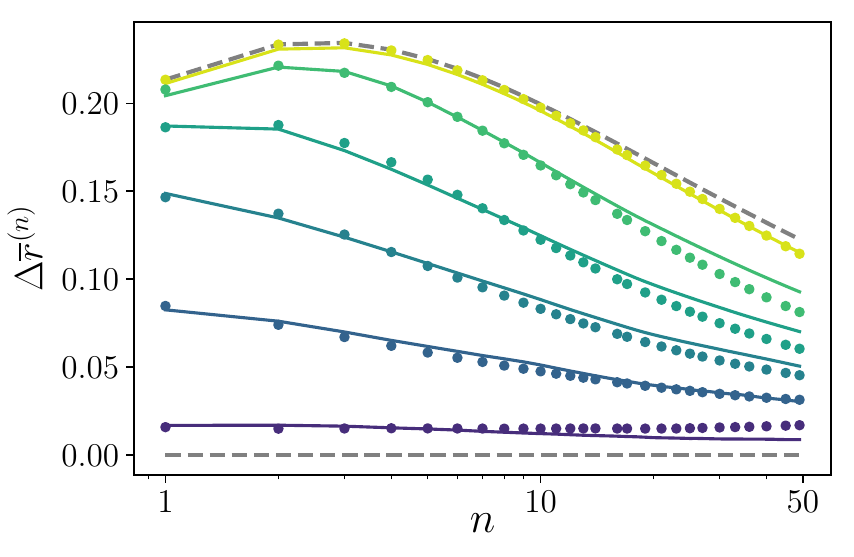}
  \caption{ { 
  The average higher order spacing ratios $\Delta \overline r^{(n)}$ as function of
  $n$ for $\epsilon=0.8, 0.5, 0.4, 0.33, 0.26, 0.15$ (from top to bottom)  for the random 
  quantum circuit are denoted by markers. Corresponding fits of $\beta$-$h$ model are denoted 
  by solid lines, the $\beta$, $h$ parameters are the same as in the main text. 
  Grey dashed lines correspond
  to $\Delta \overline r^{(n)}$ for GUE and PS respectively.
 }}\label{avrflo}
\end{figure}

Distributions of higher order spacing ratios \eqref{higher} calculated for eigenphases
$\theta_j$  are shown in Fig.~\ref{fig: higher_spacingsFlo}. {The $\beta$-$h$ model with level repulsion exponent $\beta \in [0,2]$ and appropriately 
 chosen range of interactions $h$ reproduces the distributions of higher order spacing ratios  $ P(r^{(n)})$, despite the broken time-reversal symmetry in the system. }
Average higher order gap ratios for the random quantum circuit  are shown in 
Fig.~\ref{avrflo}. The $\beta$-$h$ model gives a good account for the spectral 
correlations reflected by $\Delta \overline r^{(n)}$.
Notably, deviations at $n\gtrsim20$ suggest also in this case
 that correlations between eigenphases of the Floquet operator $W$ in the crossover regime 
 are weaker than correlations predicted by the $\beta$-$h$ model.
\begin{figure}
 \includegraphics[width=0.85\linewidth]{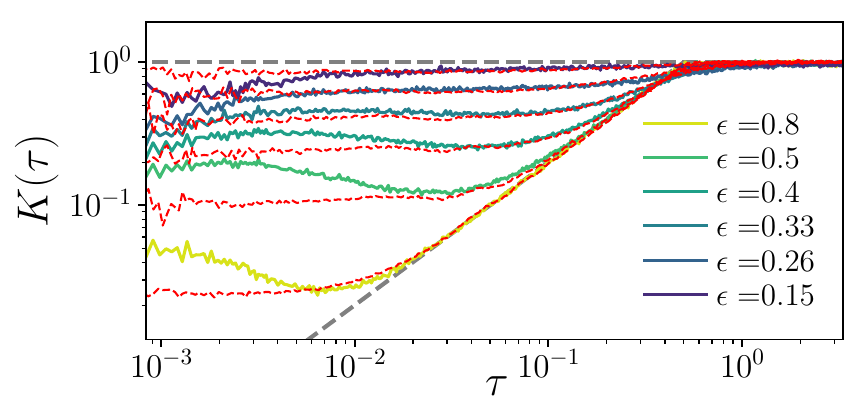}\vspace{-0.4cm}
  \caption{ { SFF for the random circuit \eqref{eq: Flo} of size $L=8$ 
  for various value of  $\epsilon$. Predictions of $\beta$-$h$ model with parameters the same as in Fig.~\ref{fig: higher_spacingsFlo}
  are denoted by red 
  dashed lines. Grey dashed lines correspond to GUE and PS.
 }} \label{fig: KtFlo} 
\end{figure}
 This suggests a similar behavior of level statistics at larger energy scales as in the case of XXZ spin chain. 
 Fig.~\ref{fig: KtFlo} shows the SFF of the considered Floquet
 operator \eqref{eq: Flo} with predictions  of $\beta$-$h$ model.
The behavior of SFF is qualitatively very similar to the case of XXZ spin chain, $K(\tau)$ follows 
the prediction for GUE down to the Thouless time $\tau_{Th}$; for smaller $\tau$, $K(\tau)$ flattens
matching  SFF of $\beta$-$h$ model. On the ergodic side of transition 
SFF of the Floquet operator increases indicating weaker correlations between eigenvalues than in the 
$\beta$-$h$ model. 
 
\section{ Discussion and outlook }
\label{sec5}

We have analyzed level statistics across the ergodic-MBL transition.
%Level statistics of the considered systems 
%of finite size depend on details of the problem \cite{Santos10}.
The proposed $\beta$-$h$ model provides a simple framework
that allows one to reproduce universal features of level statistics
of disordered interacting quantum many-body systems.
%We have demonstrated that it 
The model captures the ergodic-MBL transition
in the XXZ spin chain. Similarly, $\beta$-$h$ model 
is able to reproduce
level statistics of disordered Bose-Hubbard models that undergo ergodic-MBL transition
\cite{Sierant17, Sierant18}. The $\beta$-$h$ model grasps also 
level statistics of the random quantum circuit across the transition between ergodic and MBL
phases in spite of broken time-reversal symmetry.
Notably,  the only feature encoded in the Floquet 
operator \eqref{eq: Flo} is the locality of gates in the circuit and as such 
the random circuit
can be regarded as a toy model of a generic disorder interacting quantum system.
All this taken together  allows us to conjecture that $\beta$-$h$ model grasps universal, robust
features of level statistics of interacting disordered quantum many-body systems, 
independently, for instance, of local conservation laws \cite{Huang19,Friedman19a}.

The transition between chaotic and integrable regimes in systems with chaotic classical counterparts 
\cite{Bohigas84,Haake}
is system specific  as it is determined by the structure of underlying classical phase space \cite{Bohigas93}.
Our analysis
with $\beta$-$h$ model indicates that  spectra of %the opposite is true for 
disordered interacting quantum many-body systems are
effectively parametrized with good accuracy only by the level 
repulsion exponent $\beta$ and 
the range of interactions between eigenvalues $h$
{suggesting} an existence of a robust 
mechanism of delocalization of LIOMs that assure integrability and PS statistics in MBL phase.
Detailed understanding of such a mechanism
%, or, equivalently of mechanism 
%of decrease of the amount of level repulsion upon insertion of disorder 
%\cite{Kjall18} 
remains an open problem.
Spectral properties of disordered interacting many-body systems resemble
 level statistics at the single particle Anderson localization transition 
 \cite{Shklovskii93,Evers08} which could be expected as MBL can be regarded
 as an Anderson localization in the Hilbert space
\cite{Luca13, Logan19, Ghosh19}.

The $\beta$-$h$ model is capable of reproducing  
distributions of higher order spacing ratios, level spacing distributions, number variance 
 of systems across ergodic-MBL transition. 
%It shows that the level statistics
%at energy scales captured by those measures are effectively dependent only on the level 
%repulsion exponent $\beta$ and 
The range of interactions between eigenvalues $h$
sets the Thouless time $\tau_{Th}$ at which the SFF deviates from the universal RMT predictions.
It would be interesting to compare this time scale to 
Thouless time extracted from matrix elements of local operators \cite{Luitz17b, Serbyn17}
or from the  return probability \cite{Schiulaz19}.
The considered $\beta$-$h$ model can be also used to probe the entanglement 
spectrum 
in MBL systems 
\cite{Geraedts16} 
or random fractonic circuits
\cite{Shriya19}
as it has been shown to hosts similar, local correlations between energy levels.
It would be interesting to relate it to the associated multifractality observed deep in the MBL phase
\cite{Mace18} or to properties of level dynamics across the ergodic-MBL transition studied recently in
\cite{Maksymov19}.

 \section{acknowledgments}

  We are most grateful to Fabien Alet for kindly sharing with us the eigenvalues for $L=22,24$ Heisenberg spin chain as well as discussions
  on subjects related to this work.
We thank Dominique Delande for discussions and a careful reading of this manuscript. 
P. S. and J. Z. acknowledge support by PL-Grid Infrastructure. This research has been supported by 
 National Science Centre (Poland) under projects  2015/19/B/ST2/01028 (P.S.), 2018/28/T/ST2/00401 
 (doctoral scholarship -- P.S.) and 2016/21/B/ST2/01086 (J.Z.). 
 
 \appendix*
 \section{Remarks on unfolding}
 \label{appen}
 
 One of the advantages of analysis of level statistics with higher
order spacing ratios  $r^{(n)}$ is that they do not require spectral unfolding, i.e.
the level density $\rho(E)$ cancels out. This is of course valid only when $n$
is such that $\rho(E_i)$ and $\rho(E_{i+2n})$ are not significantly different which 
seems to be a plausible assumption when dimension of Hilbert space is larger than few thousands.

The calculation of SFF of XXZ spin chain requires application of spectral unfolding. To this end we consider 
$40000$ of eigenvalues from the center of spectrum and  fit 
the level staircase function \cite{Haake} with a polynomial of degree $10$. To calculate $K(\tau)$ 
we use $g(E)\propto \exp\left(-(E-\bar E)^2/(0.18 \Delta E^2)\right)$ (following \cite{Suntais19}) where $\bar E$ is average of the ground state and 
highest excited state energies and $\Delta E$ is standard deviation of energy in given spectrum.

In order to obtain level spacing distribution and the number variance of XXZ spin chain we consider $500$
eigenvalues from the middle of the spectrum and we perform unfolding by fitting the level staircase function 
with a third order polynomial. 

Eigenphases $\theta_j$ of the random quantum Haar-measured circuit 
are distributed uniformly in interval $[0,2\pi]$, hence no unfolding is required and 
SFF can be calculated directly from $K( t )=
\left \langle \mathrm{Tr}[W^{t}]\mathrm{Tr}[(W^{\dag})^{t}]   \right \rangle$.

%\bibliographystyle{apsrev}
%\bibliography{../../19Speckle/Draft/ref_20} 

%\input{v7.bbl}
%\include{supplement}

\end{document}